\documentclass[12pt,a4paper]{article}
\usepackage[truedimen,margin=30mm]{geometry} 

\usepackage{mathrsfs}
\usepackage{amssymb}
\usepackage{amsmath}
\usepackage{ascmac}
\usepackage{amsthm}
\usepackage[dvips]{graphicx}
\usepackage{natbib}
\usepackage{setspace}
\usepackage{times}
\usepackage{color}
\usepackage{comment}
\usepackage{algorithm}
\usepackage{algpseudocode}
\usepackage{placeins}

\usepackage{titlesec}
\titleformat*{\section}{\large\bfseries}
\titleformat*{\subsection}{\it}

\title{{\bf Fast Calibration of Generalized Posteriors \\ via Dirichlet Reweighting}\footnote{\today}}

\date{}

\begin{document}

\maketitle
\doublespacing

\vspace{-1.5cm}
\begin{center}
{\large Shonosuke Sugasawa}

\medskip

\medskip
\noindent
{\large Faculty of Economics, Keio University
}
\end{center}

\vspace{0.3cm}
\begin{center}
{\bf \large Abstract}
\end{center}

\vspace{-0cm}
While generalized Bayesian inference offers a principled approach to uncertainty quantification under general loss functions, its practical performance depends critically on the choice of the learning rate. Generalized posterior calibration, which selects the learning rate so that the resulting credible sets attain nominal frequentist coverage, is computationally intensive because it relies on repeated posterior sampling across bootstrap datasets and candidate learning rates. To overcome this practical difficulty, we propose a fast calibration method that combines Dirichlet-weighted bootstrap perturbations with importance sampling to estimate coverage while reducing the need for repeated posterior sampling. We further extend the framework to hierarchical models, where group-specific learning rates are calibrated and stabilized via shrinkage. Illustrative examples using quantile regression, support vector machines, and hierarchical Huber regression demonstrate that the proposed method substantially reduces computational cost while maintaining accurate calibration.


\newpage 
\section{Introduction}

Bayesian inference provides a principled framework for uncertainty quantification
through posterior distributions. 
In many modern applications, however, the assumed statistical model is inevitably misspecified, and standard Bayesian credible sets may then fail to attain nominal frequentist coverage, often yielding overly optimistic uncertainty quantification. 
This has motivated a growing literature on generalized Bayesian inference and Gibbs posteriors \citep{jiang2008gibbs,bissiri2016general}, in which the likelihood is replaced by a general loss function to enhance robustness against misspecification
\citep{grunwald2012safe, grunwald2020inconsistency, miller2021asymptotic, nott2024bayesian}.
In these approaches, the learning rate or temperature parameter plays a central role in uncertainty quantification: large values typically lead to undercoverage, whereas small values yield overly conservative inference. 
Reliable calibration of the learning rate is therefore essential for valid uncertainty quantification under generalized posteriors.

A variety of data-driven methods have been proposed for selecting the learning rate.
One influential line of work is information matching, which chooses the learning rate so
that an asymptotic information (or covariance) identity holding under a correctly
specified model is restored under misspecification \citep{holmes2017assigning,lyddon2019general,tamano2025location,onizuka2026learning}. 
While these methods are computationally light, they are driven primarily by large-sample
frequentist quantities and do not directly reflect the contribution of the prior to
the posterior.
Other calibration strategies have also been proposed \citep[e.g.][]{matsubara2024generalized}, but they are typically applicable to specific cases. 
A complementary approach is to calibrate the posterior directly through the
frequentist coverage of its credible sets. \citet{syring2019calibrating} introduced
generalized posterior calibration (GPC), which selects the learning rate so that the resulting credible sets attain approximately nominal coverage, with the unknown
coverage curve approximated by the nonparametric bootstrap. 
In a systematic empirical comparison, \citet{wu2023comparison} found that bootstrap-based GPC tends to deliver competitive and often superior coverage under model misspecification relative to information-matching alternatives. 
Despite its strong empirical performance, the practical use of GPC is limited by substantial computational cost, as it requires repeated Markov chain Monte Carlo (MCMC) sampling for each bootstrap dataset and each candidate learning rate. 
While a computationally efficient implementation was developed by \citet{tanaka2024generalized}, it still relies on resampling-based bootstrap replicates, which remain a dominant source of overhead.

To overcome this bottleneck, we propose an efficient calibration framework that substantially accelerates GPC by rethinking how bootstrap perturbations are incorporated. 
Our key observation is that calibration depends primarily on posterior tail probabilities and quantiles under perturbed datasets, rather than on full posterior computation. 
Moreover, under suitable regularity conditions, Dirichlet reweighting of the loss contributions reproduces the same first-order sampling fluctuations as ordinary nonparametric bootstrap resampling for regular estimators \citep{praestgaard1993exchangeably}.
This use of random weights is closely related to the Bayesian bootstrap and weighted likelihood and loss-likelihood bootstrap methods \citep{rubin1981bayesian, newton1994approximate, lyddon2019general}.
Building on this idea, we develop a \emph{Dirichlet reweighting} (DRW) approach to generalized posterior calibration.
By combining importance reweighting with reference posterior draws, the method estimates posterior tail probabilities and bootstrap coverage while substantially reducing repeated MCMC sampling; additional MCMC runs are triggered only when the effective sample size of the reweighted samples becomes too small. 
We further use this fast algorithm to calibrate group-specific learning rates in generalized posteriors for clustered data, where a shrinkage step stabilizes the group-specific estimates.

Our contributions are summarized as follows.
First, we recast bootstrap-based generalized posterior calibration as a Dirichlet reweighting problem and show that the coverage curve can be evaluated by importance reweighting of a single set of reference posterior draws, with new MCMC runs performed only when the effective sample size deteriorates. This substantially reduces the repeated MCMC sampling that dominates the cost of existing calibration methods.
Second, we extend the framework to hierarchical models by calibrating group-specific learning rates and coupling them with a shrinkage step that stabilizes the estimates when within-group sample sizes are small.
Third, through experiments on quantile regression, support vector machines, and hierarchical Huber regression, we demonstrate that the proposed method matches the coverage accuracy of existing bootstrap-based calibration at a small fraction of the computational cost.

This paper is organized as follows.
Section~\ref{sec:DRW} develops the proposed Dirichlet reweighting framework and its calibration algorithm.
Section~\ref{sec:DRW-hie} extends it to hierarchical models with group-specific learning rates.
Section~\ref{sec:illustration} presents numerical illustrations with a single learning rate, and Section~\ref{sec:Huber} considers hierarchical Huber regression.
Finally, Section~\ref{sec:conc} gives concluding remarks.

\section{Fast Posterior Calibration via Dirichlet Reweighting}\label{sec:DRW}

\subsection{Generalized posterior and learning rate}

Let $\mathcal D=\{y_1,\ldots,y_n\}$ denote the observed data and let
$\ell(\theta;y)$ be a loss function measuring the discrepancy between the model
indexed by $\theta$ and an observation $y$.
We define the empirical risk as
$R_n(\theta)=n^{-1}\sum_{i=1}^n \ell(\theta;y_i)$.
For a learning rate $\eta>0$, the corresponding generalized posterior \citep[e.g.][]{bissiri2016general} is given by
\begin{equation}\label{eq:general-pos}
\pi_\eta(\theta | \mathcal D)\propto \exp\{-n\eta R_n(\theta)\}\,p(\theta),
\end{equation}
where $p(\theta)$ denotes a prior distribution.
Let $C_{\alpha,\eta}(\mathcal D)$ be a $(1-\alpha)$ credible set constructed from
$\pi_\eta(\theta | \mathcal D)$.

The goal of calibration is to choose the learning rate $\eta$ so that the resulting
credible set achieves nominal frequentist coverage, namely
$\mathbb{P}(\theta^\dagger\in C_{\alpha,\eta}(\mathcal D))=1-\alpha$,
where $\theta^\dagger$ denotes an appropriate target parameter.
Under model misspecification, $\theta^\dagger$ is typically interpreted as the
risk minimizer associated with the population counterpart of $\ell(\theta;y)$.

Since the true data generating distribution is unknown, this coverage probability
cannot be evaluated directly.
\citet{syring2019calibrating} proposed to approximate it using the nonparametric
bootstrap.
Let $\mathcal D^\ast$ denote a bootstrap sample drawn with replacement from
$\mathcal D$.
Then the calibration condition is approximated by
$$
\mathbb{E}_{\mathcal D^\ast}
\big[
I\{\widehat\theta_\eta\in C_{\alpha,\eta}(\mathcal D^\ast)\}
\big]
\approx 1-\alpha,
$$
where $\widehat\theta_\eta$ is a point estimate computed from the original data,
such as the posterior mean under $\pi_\eta(\theta | \mathcal D)$.
In the next subsection, we show that this bootstrap-based calibration problem can be
reformulated in terms of weighted empirical risks, leading to an efficient bootstrap representation based on continuous random weights.

\subsection{Dirichlet reweighted calibration}\label{sec:DRW-algo}

Instead of the nonparametric bootstrap used in previous work \citep[e.g.][]{syring2019calibrating}, we consider a Dirichlet reweighting approach that facilitates learning-rate calibration.
In this approach, data perturbations are represented by random weights rather than by resampling observations.  
Let $\mathcal D^\ast(w)=\{(y_1,w_1),\ldots,(y_n,w_n)\}$ denote a weighted dataset that shares the same support as $\mathcal D$, where the weight vector $w=(w_1,\ldots,w_n)\sim n\cdot{\rm Dir}(1,\ldots,1)$ satisfies $\sum_{i=1}^n w_i=n$.
The weighted empirical risk associated with $\mathcal D^\ast(w)$ is defined by $R_n^\ast(\theta;w)=n^{-1}\sum_{i=1}^n w_i\,\ell(\theta;y_i)$, and the corresponding generalized posterior is
$$
\pi_\eta(\theta | \mathcal D^\ast(w))
\propto \exp\{-n\eta R_n^\ast(\theta;w)\}\,p(\theta).
$$

Let $\eta_0>0$ be a reference learning rate and let
$\theta^{(m)}\sim\pi_{\eta_0}(\theta | \mathcal D)$,
$m=1,\ldots,M$, denote posterior samples from the reference generalized posterior.
Then the density ratio between $\pi_\eta(\theta | \mathcal D^\ast(w))$ and
$\pi_{\eta_0}(\theta | \mathcal D)$ is given by
$$
r(\theta;\eta,w)
\equiv \frac{\pi_\eta(\theta | \mathcal D^\ast(w))}
{\pi_{\eta_0}(\theta | \mathcal D)}
\propto 
\exp\left\{ -\sum_{i=1}^n (\eta w_i- \eta_0) \ell(\theta; y_i)\right\}.
$$
This ratio provides importance weights for approximating expectations under the perturbed posterior indexed by $(\eta,w)$ using draws obtained under $(\eta_0,\mathcal D)$.  
For any scalar functional $g(\theta)$, the corresponding posterior distribution function under
$(\eta,w)$ can be written as
$$
F_{\eta, w}(x)\equiv \mathbb{P}(g(\theta)\leq x  |  \mathcal{D}^{\ast}(w))
= \frac{ \mathbb{E}_{\pi_{\eta_0}(\theta | \mathcal D)} \left[ I\{g(\theta)\le x\}\,r(\theta;\eta,w) \right] }{ \mathbb{E}_{\pi_{\eta_0}(\theta | \mathcal D)} \left[ r(\theta;\eta,w) \right] },
$$
where $\mathbb{E}_{\pi_{\eta_0}(\theta | \mathcal D)}$ denotes the expectation with respect to the reference posterior. 
Averaging over bootstrap perturbations yields the marginal perturbed posterior probability $\mathbb{E}_w[\mathbb{P}(g(\theta)\leq x  |  \mathcal{D}^{\ast}(w))]$.
Since the perturbed posteriors are obtained by reweighting a fixed set of reference
draws, the reliability of this importance approximation is governed by the effective
sample size (ESS). 
For the original (unweighted) data, $w^{\ast}=(1,\ldots,1)$, we use the unnormalized importance weights $r_m^{\ast}=\exp\{-n(\eta-\eta_0)R_n(\theta^{(m)})\}$, which are proportional to $r(\theta^{(m)};\eta,w^{\ast})$. Their effective sample size is defined as 
\begin{equation}\label{eq:ESS}
\mathrm{ESS}(\eta)
=\frac{\Big(\sum_{m=1}^M r_m^{\ast}\Big)^2}{\sum_{m=1}^M (r_m^{\ast})^2}
\ \in\ [1,M].
\end{equation}
This diagnostic monitors the effect of changing the learning rate for the unweighted data.
It does not directly assess weight concentration under individual bootstrap perturbations.
A small $\mathrm{ESS}(\eta)$ indicates that the reference draws obtained under $\eta_0$ poorly represent the target posterior under $\eta$; in that case the reference learning rate is refreshed by re-running MCMC at the current $\eta$.

Let $\widehat g_\eta=\mathbb{E}_{\pi_\eta(\theta\mid\mathcal D)}\{g(\theta)\}$ denote the posterior mean of the functional based on the original data, which can be estimated by importance reweighting of the reference posterior draws.
Let $C_{\alpha,\eta}(\mathcal D^{\ast}(w))$ be a $(1-\alpha)$ credible interval for $g(\theta)$ constructed from $\pi_\eta(\theta | \mathcal D^{\ast}(w))$, in the form $C_{\alpha,\eta}(\mathcal D^{\ast}(w))=\{x:\alpha/2 \le F_{\eta,w}(x)\le 1-\alpha/2\}$.
The coverage probability under learning rate $\eta$ is then defined as
\begin{equation}\label{eq:CP}
{\rm CP}_\eta
\equiv \mathbb{E}_w\left[ I\{\widehat{g}_\eta \in C_{\alpha,\eta}(\mathcal D^\ast(w))\}\right]
=\mathbb{E}_w\left[ I\left\{\frac{\alpha}{2}\le F_{\eta, w}(\widehat{g}_\eta) \le 1-\frac{\alpha}{2}
\right\}\right],
\end{equation}
where the expectation $\mathbb{E}_w$ is taken with respect to $w\sim n\cdot{\rm Dir}(1,\ldots,1)$.
This quantity measures the self-consistency of posterior tail probabilities under
Dirichlet reweighting perturbations and is used to calibrate the learning rate $\eta$.

Algorithm~\ref{alg:drw-calibration} estimates the bootstrap coverage probability in \eqref{eq:CP} using reference posterior draws.

\begin{algorithm}[htb!]
\caption{Dirichlet reweighted calibration}
\label{alg:drw-calibration}
\begin{algorithmic}[1]
\Require Candidate grid $\mathcal G$, coverage level $1-\alpha$,
         posterior sample size $M$, bootstrap size $B$, ESS threshold $\kappa$, and stopping tolerance $\varepsilon>0$
\State Set $\eta_0=\min\mathcal G$ and generate reference posterior draws
       $\theta^{(1)},\ldots,\theta^{(M)}$ by MCMC at $\eta_0$.
\For{$\eta\in\mathcal G$ in increasing order}
    \If{$\mathrm{ESS}(\eta)<\kappa$}
        \State Generate new posterior draws by MCMC at $\eta$ and set $\eta_0=\eta$.
    \EndIf
    \State Compute the unweighted-data posterior mean, with $w^\ast=\mathbf 1$,
    $$
    \widehat g_\eta
    =
    \frac{\sum_{m=1}^M r(\theta^{(m)};\eta,w^\ast)g(\theta^{(m)})}
    {\sum_{m=1}^M r(\theta^{(m)};\eta,w^\ast)}.
    $$
    \For{$b=1,\ldots,B$}
        \State Generate $w^{(b)}\sim n\,{\rm Dir}(1,\ldots,1)$ and compute
        $$
        \widehat{F}_{\eta,w}^{(b)}(\widehat g_\eta)
        =
        \frac{\sum_{m=1}^M r(\theta^{(m)};\eta,w^{(b)})
        I\{g(\theta^{(m)})\leq\widehat g_\eta\}}
        {\sum_{m=1}^M r(\theta^{(m)};\eta,w^{(b)})}.
        $$
    \EndFor
    \State Compute the coverage estimate
    $$
    \widehat{\rm CP}_\eta
    =\frac1B\sum_{b=1}^B I\left\{\frac{\alpha}{2}\leq \widehat{F}_{\eta,w}^{(b)}(\widehat g_\eta) \leq1-\frac{\alpha}{2}\right\}.
    $$
    \State \textbf{if} $\widehat{\rm CP}_\eta<1-\alpha-\varepsilon$ \textbf{then break}
\EndFor
\State Select the evaluated learning rate minimizing $|\widehat{\rm CP}_\eta-(1-\alpha)|$.
\end{algorithmic}
\end{algorithm}

The proposed procedure is computationally efficient, as it avoids MCMC sampling except when the effective sample size becomes too small.
In particular, once posterior samples are obtained from the reference generalized posterior
$\pi_{\eta_0}(\theta | \mathcal D)$, all subsequent bootstrap perturbations and learning
rate adjustments are handled through importance reweighting.
Additional MCMC runs are required only when the ESS falls below a prespecified threshold, at which point the reference learning rate $\eta_0$ is updated.
As a result, the proposed method is substantially faster than the original GPC algorithm of \citet{syring2019calibrating}, which requires independent MCMC sampling for each bootstrap replicate and each candidate learning rate.

Based on the algorithm described above, the optimal learning rate $\eta$ can be selected by solving the calibration equation ${\rm CP}_\eta=1-\alpha$.
In practice, we evaluate the estimated coverage $\widehat{\rm CP}_\eta$ along an increasing grid of candidate learning rates.
Since larger learning rates generally produce narrower credible intervals, the estimated coverage is expected to decrease, although it need not be monotone in finite samples.
Specifically, for a prespecified tolerance $\varepsilon>0$, we stop when $\widehat{\rm CP}_\eta<1-\alpha-\varepsilon$ or when the candidate grid is exhausted.
Let $\mathcal G_{\rm eval}$ denote the set of evaluated candidates.
We select $\widehat\eta\in\arg\min_{\eta\in\mathcal G_{\rm eval}}|\widehat{\rm CP}_\eta-(1-\alpha)|$, that is, the value of $\eta$ whose estimated coverage probability is closest to the nominal level among the evaluated candidates.
This sequential strategy further reduces computational cost by avoiding unnecessary evaluations at overly large learning rates while retaining accurate calibration.
In practice, we initialize the reference learning rate $\eta_0$ at the smallest value of the candidate grid and set the ESS threshold $\kappa$ to a moderate fraction of the reference sample size $M$ (e.g.\ $\kappa=M/2$), so that MCMC is refreshed only when the reweighted draws become noticeably degenerate; the results were not sensitive to this choice in our experiments.

When the parameter $\theta=(\theta_1,\ldots,\theta_p)$ is $p$-dimensional, we apply the above construction component-wise.
Specifically, for each coordinate $k=1,\ldots,p$, we consider the scalar functional $g_k(\theta)=\theta_k$ and define the corresponding coverage probability ${\rm CP}_{\eta,k}$ based on the $(1-\alpha)$ credible interval for $\theta_k$ constructed from $\pi_\eta(\theta | \mathcal D^\ast(w))$.
That is,
$$
{\rm CP}_{\eta,k}=\mathbb{E}_w\left[ I\{\widehat\theta_{\eta,k}\in C_{\alpha,\eta,k}(\mathcal D^\ast(w))\}\right],
$$
where $\widehat\theta_{\eta,k}$ denotes the $k$th component of the point estimate
$\widehat\theta_\eta$, and $C_{\alpha,\eta,k}(\mathcal D^\ast(w))$ is the marginal
credible interval for $\theta_k$.
We then define the overall coverage probability as the average of the component-wise
coverages, ${\rm CP}_\eta=p^{-1}\sum_{k=1}^p {\rm CP}_{\eta,k}$.
This component-wise aggregation provides a simple and scalable criterion for calibrating average marginal coverage.
It is particularly attractive in high-dimensional settings, where constructing and evaluating multivariate credible regions can be computationally demanding.
However, calibration of average marginal coverage does not generally ensure nominal coverage for every component or simultaneous coverage.

\subsection{Heuristic justification}\label{sec:theory}

This section provides a heuristic justification for the proposed calibration criterion.
Let $\theta^*=\arg\min_{\theta\in\Theta}R(\theta)$ be the unique interior minimizer of the population risk $R(\theta)=\mathbb{E}_P\{\ell(\theta;Y)\}$.
Under standard regularity conditions for M-estimation, the empirical risk minimizer $\hat\theta=\arg\min_{\theta\in\Theta}R_n(\theta)$ has an approximately normal sampling distribution with mean $\theta^*$ and covariance $n^{-1}J^{-1}VJ^{-1}$, where $J=\nabla^2R(\theta^*)$ is positive definite and $V=\operatorname{Var}_P\{\nabla_\theta\ell(\theta^*;Y)\}$.
For each fixed learning rate $\eta>0$, suppose also that the generalized posterior is approximately normal with center $\hat\theta$ and covariance $(n\eta)^{-1}J^{-1}$.
Thus, posterior uncertainty need not match sampling variability, and the learning rate provides a way to adjust the posterior spread.

For a scalar functional $g(\theta)$ differentiable at $\theta^*$ with nonzero gradient and positive asymptotic sampling variance, consider the frequentist coverage
\[
c_g(\eta)=\mathbb{P}\{g(\theta^*)\in C_{\eta,\alpha}(Y_{1:n})\},
\]
where $C_{\eta,\alpha}(Y_{1:n})$ is an equal-tailed credible interval for $g(\theta)$ with posterior probability $1-\alpha$.
Under the normal approximation, its width scales as $(n\eta)^{-1/2}$.
Direct evaluation of $c_g(\eta)$ is generally infeasible because the data generating distribution and $g(\theta^*)$ are unknown.
The Dirichlet weights used in Section~\ref{sec:DRW-algo} correspond to the Bayesian bootstrap \citep{rubin1981bayesian}.
Under suitable regularity conditions, Dirichlet reweighting and ordinary nonparametric resampling reproduce the same first-order sampling fluctuations for regular estimators \citep{praestgaard1993exchangeably}.
In particular, the conditional fluctuations of the Dirichlet-weighted loss minimizer around $\hat\theta$ have the same limiting sandwich covariance as the sampling fluctuations of $\hat\theta$ around $\theta^*$ \citep{lyddon2019general}.
Thus, although the two bootstrap schemes differ in finite samples, they provide similar approximations to the sampling distribution relevant to calibration.

This connection motivates using
\[
\mathrm{CP}_{g,\eta}
=\mathbb{E}_w\left[ I\{g(\hat\theta_\eta)\in C_{\eta,\alpha}(Y_{1:n};w)\} \right]
\]
as a proxy for $c_g(\eta)$, where the expectation is conditional on the observed data, $\hat\theta_\eta$ is the original-data point estimate used in calibration, and $C_{\eta,\alpha}(Y_{1:n};w)$ is the corresponding interval under the weighted generalized posterior.
For this interpretation, $g(\hat\theta_\eta)$ is assumed to be asymptotically equivalent to $g(\hat\theta)$ on the sampling fluctuation scale.
Suppose also that the weighted generalized posterior admits a normal approximation centered at the weighted empirical risk minimizer, with the same leading covariance as the original generalized posterior.
Under these approximations, the proposed algorithm seeks a learning rate for which the bootstrap coverage is close to $1-\alpha$, following the calibration principle of \cite{syring2019calibrating}.
For a fixed number of components $p$, the same reasoning applies to each component by taking $g(\theta)=\theta_k$.
The preceding approximations suggest ${\rm CP}_\eta\approx p^{-1}\sum_{k=1}^p\mathbb{P}\{\theta_k^*\in C_{\eta,\alpha,k}(Y_{1:n})\}$, where $\theta_k^*$ is the $k$th component of $\theta^*$ and $C_{\eta,\alpha,k}(Y_{1:n})$ is the marginal equal-tailed credible interval for $\theta_k$ with posterior probability $1-\alpha$.
Thus, matching ${\rm CP}_\eta$ to $1-\alpha$ aims to calibrate average marginal coverage; it does not generally calibrate every component individually or ensure simultaneous coverage.
This discussion provides a heuristic justification for the calibration criterion at fixed learning rates, rather than a formal coverage guarantee for the data-dependent selected rate.


\section{Calibrating Generalized Posteriors in Hierarchical Models}
\label{sec:DRW-hie}

\subsection{Generalized posteriors in hierarchical models}

We next discuss an extension of the proposed calibration method to hierarchical models.
Suppose that the observed data are divided into $m$ clusters.
For cluster $i=1,\ldots,m$, let $\mathcal{D}_i=\{y_{ij},\ j=1,\ldots,n_i\}$ denote the group-specific data, where $n_i$ is the within-cluster sample size.
We consider a group-specific parameter $\theta_i$ and the group loss $L_i(\theta_i;\mathcal{D}_i)=\sum_{j=1}^{n_i}\ell(\theta_i;y_{ij})$, where $\ell(\theta_i;y_{ij})$ is the loss contribution from the $j$th observation in the $i$th cluster.
A natural hierarchical generalized posterior is given by
$$
p(\theta_1,\ldots,\theta_m,\phi\mid\mathcal{D})
\propto
\exp\left\{-\sum_{i=1}^m\eta_iL_i(\theta_i)\right\}
\prod_{i=1}^m p(\theta_i;\phi)p(\phi),
$$
where $\mathcal{D}=\bigcup_{i=1}^m\mathcal{D}_i$, $L_i(\theta_i)=L_i(\theta_i;\mathcal{D}_i)$, $p(\theta_i;\phi)$ is a hierarchical prior or random-effects distribution for the group-specific parameter, and $\phi$ is a hyperparameter.
The prior $p(\theta_i;\phi)$ induces borrowing of information across clusters.
The learning rate $\eta_i$ controls the contribution of the group-specific loss $L_i(\theta_i;\mathcal{D}_i)$ to the posterior.
Since the degree of misspecification and the curvature of the loss may differ across clusters, it is natural to allow group-specific learning rates $\eta_1,\ldots,\eta_m$.

The main difficulty is that the learning rate and the hierarchical prior both affect posterior uncertainty.
In particular, changing $\eta_i$ not only changes the posterior concentration for $\theta_i$, but also changes the relative strength of borrowing from the hierarchical prior.
To address this issue, we first calibrate a common learning rate and fit the corresponding hierarchical model.
We then use the resulting posterior distribution of $\phi$ to inform group-specific calibration and stabilize the calibrated learning rates by shrinking them toward the common value.
Finally, we use these learning rates in the full hierarchical posterior.

\subsection{Calibration of a common learning rate}\label{sec:common-calibration}

We first calibrate a common learning rate within the full hierarchical model.
Write $\theta=(\theta_1,\ldots,\theta_m)$ and $L_i(\theta_i)=\sum_{j=1}^{n_i}\ell(\theta_i;y_{ij})$.
For a common rate $\eta>0$, the joint generalized posterior is
$$
q_\eta(\theta,\phi\mid\mathcal D)
\propto
\exp\left\{-\eta\sum_{i=1}^m L_i(\theta_i)\right\}
\prod_{i=1}^m p(\theta_i;\phi)p(\phi).
$$
Thus, common calibration accounts for information sharing across groups and uncertainty in the hyperparameter $\phi$.
It does not require fitting each group separately under a weakly informative prior.

For each bootstrap replicate $b=1,\ldots,B$, independently generate $w_i^{(b)}\sim n_i{\rm Dir}(1,\ldots,1)$ across groups and write $w^{(b)}=(w_1^{(b)},\ldots,w_m^{(b)})$.
One bootstrap replicate therefore perturbs observations in all groups while preserving the total weight $n_i$ within each group.
The resulting posterior is
$$
q_{\eta,w^{(b)}}(\theta,\phi\mid\mathcal D)
\propto
\exp\left\{-\eta\sum_{i=1}^m\sum_{j=1}^{n_i}
w_{ij}^{(b)}\ell(\theta_i;y_{ij})\right\}
\prod_{i=1}^m p(\theta_i;\phi)p(\phi).
$$
These perturbations operate within groups; they do not resample entire groups or reproduce residual dependence within a group.

Let $(\theta^{(s)},\phi^{(s)})$, $s=1,\ldots,S$, be reference draws from $q_{\eta_0}$.
For a candidate $\eta$, define the normalized importance weights
$$
a_s(\eta,w)=\frac{r_s(\eta,w)}{\sum_{t=1}^S r_t(\eta,w)},
\qquad
r_s(\eta,w)=\exp\left\{-\sum_{i=1}^m\sum_{j=1}^{n_i}
(\eta w_{ij}-\eta_0)\ell(\theta_i^{(s)};y_{ij})\right\}.
$$
The choice $w=\mathbf 1$ gives the unweighted-data posterior at the candidate rate.
Although calibration concerns marginal distributions of group-specific parameters, the importance weights involve losses from all groups because the reference and target are joint hierarchical posteriors.
The hierarchical prior cancels in the density ratio, but its information sharing and hyperparameter uncertainty remain represented by the joint draws.

For each group, let $g_{ik}(\theta_i)$, $k=1,\ldots,d_i$, be the scalar functionals to be calibrated; in our implementation these are the regression coefficients.
We compute the unweighted-data posterior mean and bootstrap posterior probability as
$$
\widehat g_{ik,\eta}
=\sum_{s=1}^S a_s(\eta,\mathbf 1)g_{ik}(\theta_i^{(s)}),
\qquad
\widehat F_{ik,\eta,w}(x)
=\sum_{s=1}^S a_s(\eta,w)I\{g_{ik}(\theta_i^{(s)})\le x\}.
$$
The estimated coverage criterion is
$$
\widehat{{\rm CP}}_i^c(\eta)
=\frac{1}{Bd_i}\sum_{b=1}^B\sum_{k=1}^{d_i}
I\left\{\frac{\alpha}{2}\le
\widehat F_{ik,\eta,w^{(b)}}(\widehat g_{ik,\eta})
\le 1-\frac{\alpha}{2}\right\}.
$$
This evaluates whether the bootstrap equal-tailed credible interval contains the posterior mean from the unweighted data.
For a specified set of calibration groups $\mathcal I_0$, we average these criteria with equal group weights, irrespective of sample size, and select
$$
\widehat\eta_c\in\mathop{\rm argmin}_{\eta\in\mathcal G_{\rm eval}}
\left|\frac{1}{|\mathcal I_0|}\sum_{i\in\mathcal I_0}
\widehat{{\rm CP}}_i^c(\eta)-(1-\alpha)\right|.
$$
By default, $\mathcal I_0=\{1,\ldots,m\}$; even when a subset is specified, all groups remain in the hierarchical posterior.
The set $\mathcal G_{\rm eval}$ contains the evaluated candidates from an increasing grid $\mathcal G$.
We stop when the estimated average coverage first falls below $1-\alpha-\varepsilon$, or when the grid is exhausted.
This stopping rule is a computational convention rather than a guarantee of monotonicity or of a global optimum over the full grid.
Average calibration does not imply group-specific calibration, since overcoverage in some groups may offset undercoverage in others.

To monitor the importance approximation across bootstrap perturbations, define ${\rm ESS}(\eta,w)=\{\sum_s a_s(\eta,w)^2\}^{-1}$ and
$$
H(\eta)=\frac1B\sum_{b=1}^B
I\{{\rm ESS}(\eta,w^{(b)})\le\kappa S\}.
$$
If $H(\eta)\ge\gamma$, we generate new joint posterior draws from the unweighted-data posterior $q_\eta$, set $\eta_0=\eta$, and recompute the posterior means and coverage using the same bootstrap weights.
Otherwise, the reference draws and reference rate are retained.
Our defaults are $\kappa=0.05$ and $\gamma=0.30$, both of which can be changed.
At most one reference refit is performed at each candidate rate; persistently low bootstrap ESS is reported as a warning.
No importance resampling or bootstrap-specific MCMC is used in this calibration step.
The ESS measures weight concentration rather than MCMC autocorrelation, and refreshing the unweighted reference does not guarantee adequate ESS for every bootstrap posterior.
Algorithm~\ref{alg:common-calibration} summarizes the common learning-rate calibration procedure.

\begin{algorithm}[htb!]
\caption{Common learning-rate calibration}
\label{alg:common-calibration}
\begin{algorithmic}[1]
\Require Increasing candidate grid $\mathcal G$, calibration groups $\mathcal I_0$, coverage level $1-\alpha$, posterior sample size $S$, bootstrap size $B$, thresholds $\kappa,\gamma$, and stopping tolerance $\varepsilon>0$.
\State Generate $w_i^{(b)}\sim n_i{\rm Dir}(1,\ldots,1)$ independently for $i=1,\ldots,m$ and $b=1,\ldots,B$; retain these weights throughout the search.
\State Set $\eta_0=\min\mathcal G$ and $\mathcal G_{\rm eval}=\varnothing$.
\State Generate joint posterior draws $(\theta^{(s)},\phi^{(s)})$, $s=1,\ldots,S$, from $q_{\eta_0}$.
\For{$\eta\in\mathcal G$ in increasing order}
    \State Compute the normalized importance weights $a_s(\eta,\mathbf 1)$ and $a_s(\eta,w^{(b)})$.
    \State Calculate ${\rm ESS}(\eta,w^{(b)})=\{\sum_{s=1}^S a_s(\eta,w^{(b)})^2\}^{-1}$ and
    $$
    H(\eta)=\frac{1}{B}\sum_{b=1}^B
    I\{{\rm ESS}(\eta,w^{(b)})\le\kappa S\}.
    $$
    \If{$H(\eta)\ge\gamma$}
        \State Generate new joint posterior draws from $q_\eta$ and set $\eta_0=\eta$.
        \State Recompute the importance weights using the retained bootstrap weights.
    \EndIf
    \State Compute the posterior means $\widehat g_{ik,\eta}$ and bootstrap posterior distribution functions $\widehat F_{ik,\eta,w^{(b)}}$ to evaluate 
    $$
    \widehat{\rm CP}_i^c(\eta)
    =\frac{1}{Bd_i}\sum_{b=1}^B\sum_{k=1}^{d_i}
    I\left\{\frac{\alpha}{2}\le
    \widehat F_{ik,\eta,w^{(b)}}(\widehat g_{ik,\eta})
    \le1-\frac{\alpha}{2}\right\},
    \qquad i\in\mathcal I_0.
    $$
    \State Record $\widehat C(\eta)=|\mathcal I_0|^{-1}\sum_{i\in\mathcal I_0}\widehat{\rm CP}_i^c(\eta)$ and add $\eta$ to $\mathcal G_{\rm eval}$.
    \State \textbf{if} $\widehat C(\eta)<1-\alpha-\varepsilon$ \textbf{then break}
\EndFor
\State 
Compute $\widehat\eta_c\in\mathop{\rm argmin}_{\eta\in\mathcal G_{\rm eval}}|\widehat C(\eta)-(1-\alpha)|$.
\end{algorithmic}
\end{algorithm}

\subsection{Group-specific calibration via shrinkage} \label{sec:group-calibration}

We fit the hierarchical model with $\eta_i=\widehat\eta_c$ for all $i$ and denote the resulting marginal posterior distribution of $\phi$ by $q_c(\phi)$.
Using this pilot posterior, we define
$$
\overline p_i(\theta_i)=\int p(\theta_i;\phi)q_c(\phi)\,d\phi,
\qquad
\pi_{i,\eta}^{c}(\theta_i\mid\mathcal D_i)
\propto \exp\{-\eta L_i(\theta_i)\}\overline p_i(\theta_i).
$$
In computation, we replace the integral by the empirical mixture $S_\phi^{-1}\sum_{s=1}^{S_\phi}p(\theta_i;\phi^{(s)})$, using paired hyperparameter draws from the pilot fit.
We retain this mixture throughout individual calibration and variance estimation, rather than substituting the posterior mean of $\phi$.
In the Huber regression implementation, local MCMC updates only $\theta_i$, evaluating the mixture prior density in the Metropolis ratio.
Thus, uncertainty within the pilot distribution of $\phi$ is propagated, while the pilot distribution itself is not re-estimated under the perturbations.
This is a data-dependent prior used for calibration, and $\pi_{i,\eta}^{c}$ need not coincide with the marginal posterior under the full hierarchy.
Consequently, a new local reference sample is generated under this mixture prior; pilot draws of $\theta_i$ are used for initialization only.

Let $\lambda_i=\log\eta_i$.
For group $i$, define $\widehat{{\rm CP}}_i(\lambda_i)$ by the same posterior-tail criterion as above, using the unweighted-data posterior mean under $\pi_{i,e^{\lambda_i}}^c$ and bootstrap posteriors proportional to
$$
\exp\left\{-e^{\lambda_i}\sum_{j=1}^{n_i}w_{ij}\ell(\theta_i;y_{ij})\right\}\overline p_i(\theta_i),
\qquad w_i\sim n_i{\rm Dir}(1,\ldots,1).
$$
We average the criterion over $g_{ik}$ when multiple functionals are considered.
With local reference draws from $\pi_{i,\eta_0}^c$, the importance weights are proportional to $\exp\{-\sum_j(\eta w_{ij}-\eta_0)\ell(\theta_i^{(s)};y_{ij})\}$ and depend only on the target group's loss.
We initialize the local reference at $\widehat\eta_c$, search the increasing candidate grid, and use the same low-bootstrap-ESS rule as in Algorithm~\ref{alg:common-calibration} to refresh the unweighted-data local reference.
The search stops when the estimated group coverage first falls below $1-\alpha-\varepsilon$, or when the grid is exhausted.
The raw calibrated log rate $\lambda_i^\ast$ is the logarithm of the evaluated rate whose coverage is closest to $1-\alpha$.

Although individual calibration accommodates group-wise heterogeneity, $\lambda_i^\ast$ can be unstable when $n_i$ is small.
This instability is not merely Monte Carlo error from a finite number of bootstrap weights: it also reflects variation in the calibrated rate as a function of the data.
We therefore estimate its variability and shrink the raw log rates toward the common calibrated value.
Let $\lambda_{i,0}$ be a locally unique solution to $\Psi_i(\lambda)={\rm CP}_i(\lambda)-(1-\alpha)=0$, with $\Psi_i'(\lambda_{i,0})\ne0$.
A first-order expansion around $\lambda_{i,0}$ suggests
$$
{\rm Var}(\lambda_i^\ast)
\approx
\frac{{\rm Var}\{\widehat{\rm CP}_i(\lambda_{i,0})\}}
{\{\Psi_i'(\lambda_{i,0})\}^2}.
$$
In computation, we replace $\lambda_{i,0}$ by $\lambda_i^\ast$ when estimating the numerator and slope.
This motivates a local variance approximation rather than an exact identity for the discrete grid-selected estimate.

We approximate the slope by $\widehat D_i=\{\widehat{{\rm CP}}_i(\lambda_i^\ast+h)-\widehat{{\rm CP}}_i(\lambda_i^\ast-h)\}/(2h)$.
Both evaluations initially use the same reference draws generated at $e^{\lambda_i^\ast}$ and the same Dirichlet weights.
If the low-ESS rule triggers a reference refit at either perturbed rate, its posterior draws are regenerated, while the bootstrap weights remain unchanged.
Thus, sharing reference draws reduces Monte Carlo variation when no such refit is needed.

We estimate the numerator by a double weighted bootstrap.
For $r=1,\ldots,R$, generate $u_i^{(r)}\sim n_i{\rm Dir}(1,\ldots,1)$ and consider the outer posterior proportional to
$$
\exp\left\{-e^{\lambda_i^\ast}\sum_j u_{ij}^{(r)}\ell(\theta_i;y_{ij})\right\}\overline p_i(\theta_i).
$$
Starting from the unweighted local reference at $e^{\lambda_i^\ast}$, compute its importance ESS.
If this ESS is at most $\kappa_{\rm out}S$, generate fresh local MCMC draws from the outer posterior, still using the same mixture prior; the default is $\kappa_{\rm out}=0.05$.
Otherwise, evaluate the outer posterior by importance sampling.
In either case, recompute the outer posterior mean to serve as the center of the inner coverage calculation.

Independently generate inner weights $w_i^{(r,b)}\sim n_i{\rm Dir}(1,\ldots,1)$ for $b=1,\ldots,B_{\rm var}$.
The inner weighted loss is $\sum_j u_{ij}^{(r)}w_{ij}^{(r,b)}\ell(\theta_i;y_{ij})$, with product weights used without renormalization.
We use this two-stage weighting scheme as a local approximation to the sampling variability of the calibration criterion.
Since the product weights do not generally sum to $n_i$, the inner perturbation also changes the total weight assigned to the loss.
For reference draws generated at rate $\eta_0$ with observation weights $v_i$, an unnormalized importance weight is
$$
r_{i,s}(\eta,t_i;\eta_0,v_i)
=\exp\left\{-\sum_j(\eta t_{ij}-\eta_0v_{ij})\ell(\theta_i^{(s)};y_{ij})\right\}.
$$
Here, the outer and inner targets have $t_i=u_i^{(r)}$ and $t_i=u_i^{(r)}w_i^{(r,b)}$, respectively, and $\eta=e^{\lambda_i^\ast}$.
The reference weights are $v_i=\mathbf 1$ without an outer refit and $v_i=u_i^{(r)}$ after one.
This distinction ensures that an outer perturbation is not counted twice.
Inner posterior probabilities are evaluated by importance sampling without additional MCMC.
Outer-refitted draws are used only within the corresponding outer replicate.

Denote the resulting inner-bootstrap coverage estimate by $\widehat{{\rm CP}}_i^{(r)}(\lambda_i^\ast)$ and its average over $r$ by $\overline{{\rm CP}}_i(\lambda_i^\ast)$.
We calculate
$$
\widehat v_i=\frac1R\sum_{r=1}^R
\{\widehat{{\rm CP}}_i^{(r)}(\lambda_i^\ast)-\overline{{\rm CP}}_i(\lambda_i^\ast)\}^2,
\qquad
\widehat\sigma_i^2
=\min\left\{v_{\max},\max\left\{v_{\min},
\frac{\widehat v_i}{\widehat D_i^2+\rho^2}\right\}\right\}.
$$
This avoids recalibrating the learning rate for every outer perturbation.
Our implementation uses $h=0.1$, $\rho=0.05$, $v_{\min}=10^{-6}$, and $v_{\max}=100$.
The pilot-informed mixture prior and residual scales are held fixed throughout this calculation.
Consequently, this conditional variance approximation does not incorporate uncertainty from re-estimating the pilot hierarchy or the common learning rate.
Finite inner-bootstrap and posterior-sampling error may also contribute to $\widehat v_i$.
Persistently low inner ESS is reported rather than triggering inner MCMC refits.

Based on the raw rates and estimated variances, we use
$$
\lambda_i^\dagger=\widehat\lambda_c+\frac{\widetilde V_0}{\widehat\sigma_i^2+\widetilde V_0}
(\lambda_i^\ast-\widehat\lambda_c),
\qquad \widehat\lambda_c=\log\widehat\eta_c,
$$
where $\widetilde V_0$ minimizes
$$
Q(V_0)=\sum_{i\in\mathcal I_1}
\left[\log(\widehat\sigma_i^2+V_0)
+\frac{(\lambda_i^\ast-\widehat\lambda_c)^2}{\widehat\sigma_i^2+V_0}\right]-\log V_0,
\qquad V_0>0.
$$
Here, $\mathcal I_1$ is the set of individually calibrated groups.
Apart from the adjustment term and an additive constant, $Q(V_0)$ is twice the negative log-likelihood of the working model $\lambda_i^\ast\sim N(\widehat\lambda_c,\widehat\sigma_i^2+V_0)$.
The term $-\log V_0$ discourages a zero dispersion estimate, following the adjusted-likelihood approach of \citet{li2010adjusted}.
The corresponding $\lambda_i^\dagger$ is an empirical Bayes posterior mean under this normal working hierarchy, with the center fixed at the common calibrated log rate.
We set $\eta_i^\ast=\exp(\lambda_i^\dagger)$; groups without individual calibration use $\eta_i^\ast=\widehat\eta_c$.
In the implementation, $\mathcal I_1=\{i:n_i>1\}$, so singleton groups retain the common rate.
Algorithm~\ref{alg:group-calibration} summarizes the group-specific calibration procedure, including double-bootstrap variance estimation, shrinkage, and the final hierarchical fit.

The common rate pools information across groups and can therefore be comparatively stable, but its average criterion may conceal group-specific undercoverage or overcoverage.
Individual calibration accommodates heterogeneity but may be unstable with small within-group samples.
The shrinkage rule balances these features: groups with larger estimated variances are shrunk more strongly, whereas more precise individual estimates retain their adjustments.
Greater dispersion of the raw log rates relative to their estimated variability tends to favor a larger $\widetilde V_0$ and less shrinkage.

The working model constructs a regularization rule rather than a fully Bayesian model for the learning rates.
It treats the common target and estimated variances as fixed and does not account for dependence among the calibrated values.
Moreover, shrinkage need not preserve the individual or average calibration equations, so its effect on coverage is evaluated empirically.
Finally, we fit the full hierarchical generalized posterior using $\eta_i=\eta_i^\ast$, jointly updating $\theta_1,\ldots,\theta_m$ and $\phi$.
This final fit uses the original hierarchical prior, not the local mixture priors, and conditions on the selected learning rates.

\begin{algorithm}[htb!]
\caption{Group-specific calibration with double bootstrap and shrinkage}
\label{alg:group-calibration}
\begin{minipage}{\linewidth}
\small
\begin{algorithmic}[1]
\Require Groups $\mathcal I_1$, grid $\mathcal G$, level $1-\alpha$, sample sizes $S,S_\phi,B,R,B_{\rm var}$, and tuning constants specified in Section~\ref{sec:group-calibration}
\State Obtain $\widehat\eta_c$ from Algorithm~\ref{alg:common-calibration} and set $\widehat\lambda_c=\log\widehat\eta_c$.
\State Fit the hierarchy at the common rate and fix $\overline p_i(\theta_i)=S_\phi^{-1}\sum_s p(\theta_i;\phi^{(s)})$ using pilot draws $\phi^{(s)}\sim q_c$.
\For{$i\in\mathcal I_1$}
    \State Initialize local reference draws from $\pi_{i,\widehat\eta_c}^c$.
    \State Search $\mathcal G$ using local coverage and the reference-refit rule of Algorithm~\ref{alg:common-calibration}, stopping at the first coverage below $1-\alpha-\varepsilon$ or when the grid is exhausted.
    \State Select $\lambda_i^\ast=\log\eta_i^{\rm raw}$, where $\eta_i^{\rm raw}$ minimizes $|\widehat{\rm CP}_i(\log\eta)-(1-\alpha)|$ over evaluated rates.
    \State Draw a local reference sample at $e^{\lambda_i^\ast}$ and estimate
    $$
    \widehat D_i=
    \frac{\widehat{\rm CP}_i(\lambda_i^\ast+h)
    -\widehat{\rm CP}_i(\lambda_i^\ast-h)}{2h},
    $$
    using shared bootstrap weights and the reference-refit rule.
    \For{$r=1,\ldots,R$}
        \State Generate outer weights $u_i^{(r)}\sim n_i{\rm Dir}(1,\ldots,1)$.
        \State Starting from the reference at $e^{\lambda_i^\ast}$, evaluate the outer posterior by importance sampling; refit by MCMC if its ESS is at most $\kappa_{\rm out}S$.
        \State Compute the outer posterior means and generate $B_{\rm var}$ independent inner Dirichlet weight vectors $w_i^{(r,b)}$.
        \State Using unnormalized product weights $u_i^{(r)}w_i^{(r,b)}$, evaluate inner posteriors by importance sampling from the current reference.
        \State Compute $\widehat{\rm CP}_i^{(r)}(\lambda_i^\ast)$ as the average inner interval coverage of the outer posterior means.
    \EndFor
    \State Compute $\widehat v_i$, the empirical variance with divisor $R$ of the outer coverage estimates, and set $\widehat\sigma_i^2=
    \min\{v_{\max},\max\{v_{\min},
    \widehat v_i/(\widehat D_i^2+\rho^2)\}\}$.
\EndFor
\State Obtain $\widetilde V_0=\arg\min_{V_0>0}Q(V_0)$ and, for $i\in\mathcal I_1$, set
$$
\lambda_i^\dagger=\widehat\lambda_c+
\frac{\widetilde V_0}{\widehat\sigma_i^2+\widetilde V_0}
(\lambda_i^\ast-\widehat\lambda_c),
\qquad \eta_i^\ast=e^{\lambda_i^\dagger}.
$$
\State Set $\eta_i^\ast=\widehat\eta_c$ for $i\notin\mathcal I_1$.
\State Fit the full hierarchy with the original hierarchical prior at $\eta_i=\eta_i^\ast$, jointly updating all group parameters and $\phi$.
\end{algorithmic}
\end{minipage}
\end{algorithm}

\FloatBarrier

\section{Illustrations: Single Learning Rate}\label{sec:illustration}

We demonstrate the proposed calibration algorithm through two examples considered by \citet{syring2019calibrating}.
In what follows, all computation times were measured on a Mac Pro equipped with a 3.5 GHz 8-core Intel Xeon W processor and 48 GB of memory, using R version 4.6.1.

\subsection{Quantile regression}
Let $y_i$ be a response variable and $x_i$ be a vector of covariates. 
For Bayesian quantile regression, we use the asymmetric Laplace working likelihood, whose posterior distribution coincides with the Gibbs posterior based on the empirical check loss. 
Specifically, letting $\rho_{\tau}(u)=u\{\tau-I(u<0)\}$, the generalized posterior distribution is given by
$$
\pi_\eta(\beta\mid\mathcal D)
\propto
\exp\left\{-\eta\sum_{i=1}^n
\rho_\tau(y_i-x_i^\top\beta)\right\}p(\beta),
$$
where $\eta>0$ is the learning rate and $\mathcal{D}=\{(x_i,y_i), \ i=1,\ldots,n\}$ is the observed data.
Posterior computation is performed using a Gibbs sampler based on data augmentation \citep{kozumi2011gibbs}.

We generate covariates $X=(1,X_1,X_2,X_3)$, where $X_j\sim N(0,1)$ independently, and set the coefficient vector in the data generating location model to $\beta_0=(0.5,1.0,-0.7,0)^\top$.
The response is generated as $y_i=x_i^\top\beta_0+\varepsilon_i$, where $\varepsilon_i$ follows a standard $t$ distribution with three degrees of freedom.
We fix the target quantile level at $\tau=0.75$ and consider sample sizes $n\in\{100,500\}$.
The corresponding conditional quantile coefficient vector is $\beta_\tau=\beta_0+t_3^{-1}(\tau)e_1$, where $e_1=(1,0,0,0)^\top$ and $t_3^{-1}$ denotes the quantile function of the standard $t_3$ distribution.
Thus, the non-intercept coefficients remain unchanged, and we focus on these coefficients when evaluating estimation accuracy and coverage.

For quantile regression, we construct the generalized posterior based on a Gaussian prior, $\beta\sim N(0, s_0^2 I_4)$ with $s_0^2=10$.
The learning rate $\eta$ is selected via coverage calibration at nominal level $1-\alpha=0.95$, using a grid $\eta\in\{0.1,0.2,\ldots,3.0\}$.
We compare three calibration strategies: the proposed DRW method, the sequential Monte Carlo (SMC) calibration method proposed by \citet{tanaka2024generalized}, and the brute-force generalized posterior calibration (GPC) algorithm based on the nonparametric bootstrap proposed by \citet{syring2019calibrating}.
Posterior computation uses $M=3000$ retained draws after $1000$ burn-in iterations.
The frequentist performance is assessed based on $R=100$ Monte Carlo replications.
We report the root mean squared error (RMSE) of the posterior mean estimator and the empirical coverage probability (CP), computed from indicators that the $95\%$ marginal credible intervals contain their corresponding true coefficients and averaged over the non-intercept components and Monte Carlo replications.
In addition, we record the total computation time required for the calibration procedure in each replication.
Table~\ref{tab:Sim-qr} reports the results. 
Across all settings, DRW yields RMSE and empirical coverage similar to those of
SMC and brute-force GPC, while considerably reducing the computation time.
The calibrated learning rates also agree closely across the three methods.

\begin{table}
\medskip
\begin{tabular}{cccccccccccc}
\hline
$(n,B)$  & \multicolumn{3}{c}{(100, 200)} && \multicolumn{3}{c}{(100, 500)} && \multicolumn{3}{c}{(500, 200)} \\
\cline{2-4}\cline{6-8}\cline{10-12}
& {\small DRW} & {\small SMC} & {\small GPC} &&   {\small DRW} & {\small SMC} & {\small GPC} &&   {\small DRW} & {\small SMC} & {\small GPC} \\
\hline
CPT & 1.1 & 15.3 & 53.5 &  & 2.7 & 37.1 & 130.7 &  & 7.9 & 50.6 & 159.5 \\
RMSE & 0.153 & 0.153 & 0.154 & & 0.158 & 0.157 & 0.157 &  & 0.071 & 0.069 & 0.071 \\
CP (\%) & 96.7 & 95.7 & 95.7 &  & 95.0 & 95.7 & 93.3 &  & 96.0 & 95.3 & 96.7 \\
$\eta$ & 1.49 & 1.56 & 1.60 &  & 1.46 & 1.56 & 1.61 &  & 1.36 & 1.47 & 1.39 \\
\hline
\end{tabular}
\caption{Computation time (CPT) in seconds, root mean squared error (RMSE) of the posterior means, coverage probability (CP) of $95\%$ marginal credible intervals, and calibrated learning rate $\eta$, averaged over 100 Monte Carlo replications.
RMSE and CP are summarized over the three non-intercept coefficients.
DRW denotes the proposed Dirichlet reweighting method, SMC denotes the sequential Monte Carlo calibration algorithm, and GPC denotes the brute-force generalized posterior calibration algorithm based on the nonparametric bootstrap.}
\label{tab:Sim-qr}
\end{table}

\subsection{Support vector machines}

We next illustrate the proposed calibration method using support vector machine (SVM) classification.
Let $y_i\in\{-1,1\}$ be a binary response and $x_i$ be a $p$-dimensional covariate vector including an intercept.
The linear SVM classifies a new observation with covariate vector $x$
according to the sign of $x^\top\beta$, where $\beta$ is the coefficient vector.
To perform Bayesian inference on $\beta$, we consider the empirical risk based on the hinge loss
$$
R_n(\beta)=\frac{2}{n}\sum_{i=1}^n \max\{0,1-y_i x_i^\top\beta\}.
$$
The corresponding population target is the coefficient vector minimizing the population counterpart of this risk.
To construct a generalized posterior, we assign independent zero-centered Laplace priors with scale parameters $\nu\sigma_k$: $p(\beta)\propto \exp(-\nu^{-1}\sum_{k=1}^p\sigma_k^{-1}|\beta_k|)$, where $\nu>0$ is fixed, $\sigma_1=1$ for the intercept, and $\sigma_k$ is the sample standard deviation of the corresponding predictor for $k=2,\ldots,p$.
The resulting generalized posterior of $\beta$ is given by
$$
\pi_\eta(\beta|\mathcal{D})
\propto \exp\left\{ -2\eta \sum_{i=1}^n \max\{0,1-y_i x_i^\top\beta\} - \frac{1}{\nu}\sum_{k=1}^p \frac{|\beta_k|}{\sigma_k} \right\},
$$
where $\mathcal{D}$ denotes the observed dataset. 
The learning rate $\eta$ determines the relative contribution of the hinge loss to the generalized posterior and, consequently, controls the concentration of the posterior distribution.
We use the data-augmentation Gibbs sampler developed by \citet{polson2011data} to generate posterior draws of $\beta$ from the above posterior distribution.

To illustrate and compare the empirical performance of the learning-rate selection methods, we employ the South African Heart Disease dataset, which consists of $n=462$ individuals, as used by \citet{syring2019calibrating} and \citet{tanaka2024generalized}.
The response is coded as $y_i=-1$ for the presence of coronary heart disease and $y_i=1$ for its absence.
We use four predictors: cumulative tobacco consumption, low-density lipoprotein cholesterol, family history of heart disease, and age.
Together with the intercept, the dimension of the coefficient vector is therefore $p=5$.
The predictors are used on their original scales, and the prior scale for coefficient $\beta_k$ is set to $\nu\sigma_k$.
We fix the prior tuning parameter at $\nu=10$.

We calibrate the learning rate to target a nominal coverage probability of $0.95$ over the grid $\eta\in\{0.05,0.06,\ldots,0.5\}$.
We compare the proposed Dirichlet reweighting (DRW) method with the sequential Monte Carlo (SMC) calibration \citep{tanaka2024generalized} and the brute-force GPC method based on the nonparametric bootstrap \citep{syring2019calibrating}.
We consider bootstrap sizes $B\in\{100,200,500\}$.
For each setting, we use $B$ Dirichlet perturbations for DRW and $B$ nonparametric bootstrap samples for SMC and GPC.
The number of reference posterior draws or particles is set to $M=1000$ after discarding the first $1000$ iterations as burn-in.
The calibrated learning rate is selected as the candidate value whose estimated average marginal coverage was closest to $0.95$ (i.e., $\alpha=0.05$), using the stopping rule described in Section~\ref{sec:DRW-algo}.
The computation times and selected learning rates for the three bootstrap sizes, $B\in\{100,200,500\}$, are reported in Table~\ref{tab:SVM}. 
The results show that DRW requires considerably less computation time than the nonparametric bootstrap-based calibration methods (SMC and GPC), and the improvement becomes more pronounced for larger $B$.
Nonetheless, the learning rate $\eta$ selected by DRW is close to the values selected by the other methods.

\begin{table}
\medskip
\begin{tabular}{cccccccccccc}
\hline
& \multicolumn{3}{c}{$B=100$} && \multicolumn{3}{c}{$B=200$} && \multicolumn{3}{c}{$B=500$} \\
\cline{2-4}\cline{6-8}\cline{10-12}
 & DRW & SMC & GPC &  & DRW & SMC & GPC &  & DRW & SMC & GPC \\
\hline
CPT & 4 & 121 & 2251 &  & 5 & 252 & 4273 &  & 12 & 631 & 11028 \\
$\eta$ & 0.29 & 0.26 & 0.26 &  & 0.26 & 0.27 & 0.26 &  & 0.25 & 0.26 & 0.25 \\
\hline
\end{tabular}
\caption{Computation times (CPT) in seconds and learning rates $\eta$ selected by Dirichlet reweighting (DRW), sequential Monte Carlo (SMC), and generalized posterior calibration (GPC) for bootstrap sizes $B\in\{100,200,500\}$ in the SVM analysis of the South African Heart Disease dataset.  }
\label{tab:SVM}
\end{table}

\begin{figure}[tb]
\centering
\includegraphics[width=\textwidth]{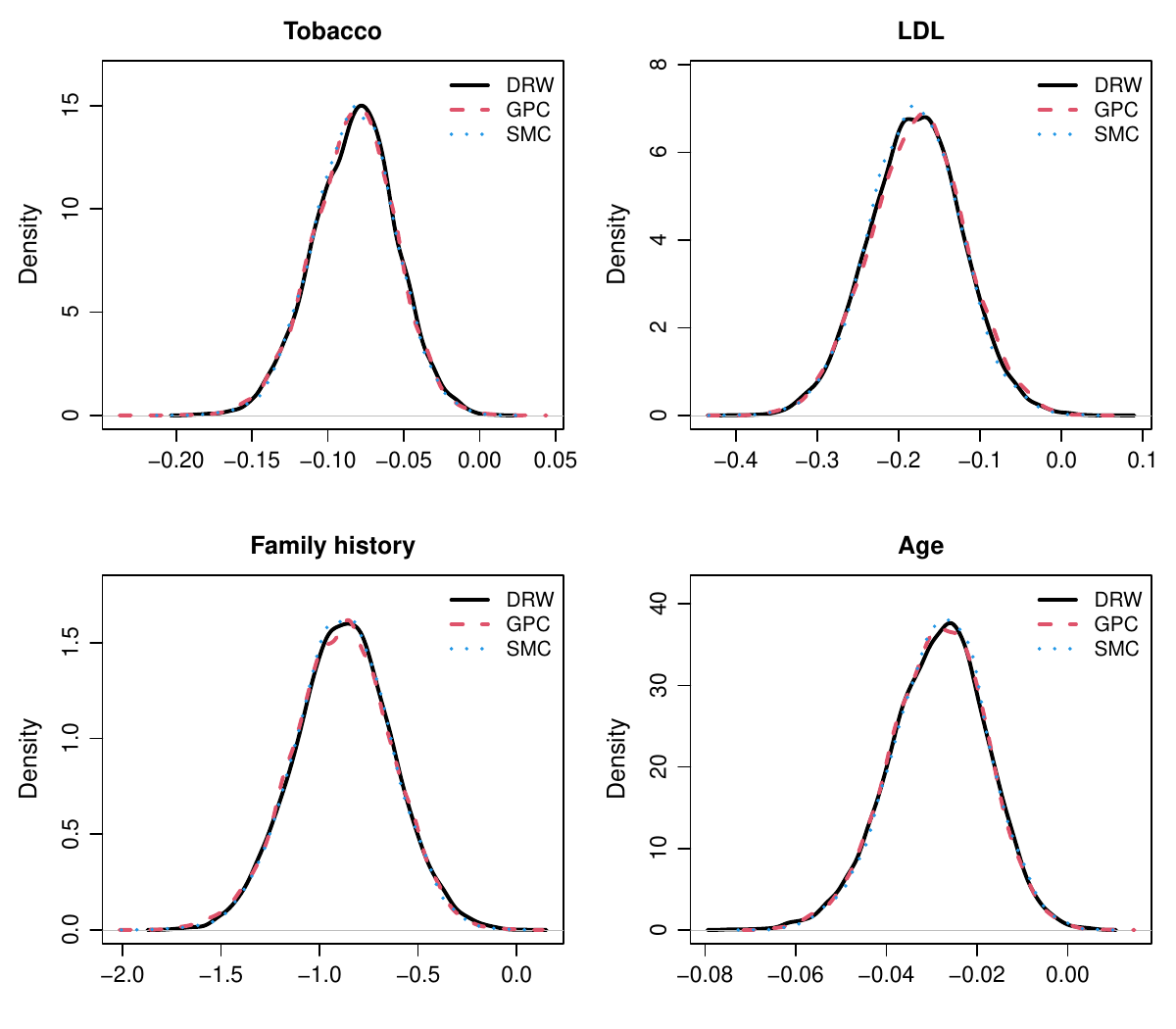}
\caption{Marginal posterior densities of the coefficients for tobacco, LDL, family history, and age in the SVM example, using the learning rates calibrated by DRW (black solid), GPC (red dashed), and SMC (blue dotted) with $B=500$.}
\label{fig:SVM-posterior}
\end{figure}

Figure~\ref{fig:SVM-posterior} compares the marginal posterior densities of the coefficients for tobacco, LDL, family history, and age under the learning rates selected by the three methods with $B=500$.
The posterior densities nearly overlap for all four coefficients, indicating that the differences in the calibrated learning rates have little effect on the resulting marginal posterior distributions in this example.
Thus, the computational savings of DRW are achieved while providing posterior inference closely comparable to that of SMC and GPC.

\section{Illustration: Hierarchical Huber Regression}\label{sec:Huber}

\subsection{Setup}
We next consider a hierarchical Bayesian formulation of classical Huber regression \citep{huber1973robust}.
Suppose that the observations are divided into multiple groups. 
For each group $i=1,\ldots,m$, we consider the linear regression model $y_{ij}=x_{ij}^{\top}\beta_i+
\varepsilon_{ij}$ for $j=1,\ldots,n_i$, where $\beta_i$ is the group-specific regression coefficient vector, $x_{ij}$ is the corresponding covariate vector and $n_i$ is the group-specific sample size.
Robust inference for $\beta_i$ under contamination is obtained using the Huber loss function
$$
\rho_{\delta}(u)
=
\begin{cases}
u^2/2, & |u|\leq\delta,\\
\delta|u|-\delta^2/2, & |u|>\delta,
\end{cases}
$$
where $\delta>0$ is a fixed tuning parameter. 
Given a group-specific learning rate $\eta_i>0$, the generalized likelihood contribution from the group is defined as
\begin{equation}\label{eq:Huber}
\exp \left\{ -\eta_i \sum_{j=1}^{n_i}
\rho_{\delta}\big(y_{ij}-x_{ij}^{\top}\beta_i\big) \right\}.
\end{equation}
To borrow information across groups, we assume the coefficient-wise hierarchical model $\beta_{ik}|\mu_k,\tau_k^2\sim N(\mu_k,\tau_k^2)$ independently for $k=1,\ldots,p$, where $\beta_i=(\beta_{i1},\ldots,\beta_{ip})^\top$.
For the global parameters $\mu_k$ and $\tau_k^2$, we assign the priors $\mu_k \sim N(m_{0k},s_{0k}^2)$ and $\tau_k^2\sim{\rm IG}(a_{0k},b_{0k})$.
The resulting model therefore combines robustness against outlying observations within each group with information sharing across groups through the hierarchical prior on $\beta_i$. 

For fixed learning rates, posterior computation is performed using a Metropolis-within-Gibbs sampler.
We update each coordinate of $\beta_i$ using a Gaussian random-walk Metropolis step, while $\mu_k$ and $\tau_k^2$ are sampled directly from their normal and inverse-gamma full conditional distributions, respectively.
Although latent-variable Gibbs sampling is available for related pseudo-Huber regression models \citep{kawakami2023approximate}, we use Metropolis updates to retain the original Huber loss in \eqref{eq:Huber}.

\subsection{Simulation experiments}\label{sec:Huber-sim}

We conducted simulation experiments to investigate the finite-sample performance of the proposed calibration and shrinkage procedures for hierarchical Huber regression.
We considered two simulation settings with $m\in\{20,60\}$ groups.
For all settings, the number of observations within each group was fixed at $n_i=15$ for $i=1,\ldots,m$.
Thus, the amount of information available within each group was kept small, even when the total number of groups increased.
In each group, we considered a regression model with an intercept and a single covariate, so that $x_{ij}=(1,z_{ij})^{\top}$, where $z_{ij}$ was generated independently from the uniform distribution on $[-1,1]$.
The true group-specific regression coefficients were generated independently from $\beta_{i1}\sim N(0,0.7^2)$ and $\beta_{i2}\sim N(1,0.5^2)$.
The response variable was then generated from $y_{ij}=x_{ij}^{\top}\beta_i+\varepsilon_{ij}$.
To introduce outlying observations, the error distribution was specified as $\varepsilon_{ij}\sim (1-\omega_i)N(0,0.5^2)+\omega_i N(0,5^2)$, where the contamination probabilities $\omega_i$ were generated from the uniform distribution on $[0,0.3]$.
For each setting, these probabilities were held fixed across Monte Carlo replications, while the regression coefficients, covariates, and responses were regenerated in each replication.
For each simulated dataset, we applied the hierarchical Huber model consisting of the generalized likelihood (\ref{eq:Huber}) and the hierarchical prior for $\beta_i$.
The tuning parameter in the Huber loss was fixed at $\delta=1.345$.
The residual scale in the loss was fixed at one for all groups and methods.
For all methods, we set $m_{0k}=0$, $s_{0k}^2=100$, $a_{0k}=1$, and $b_{0k}=1$ for each coefficient $k$.
For calibrating the learning rates, we used the grid $\{0.2,0.4,\ldots,4.0\}$.
The nominal credibility level was set to $95\%$, corresponding to $\alpha=0.05$.

We first calibrated a common learning rate using the average bootstrap coverage across groups and coefficients.
We then fitted the full hierarchy at this rate and used posterior draws of the hyperparameters to construct the posterior-averaged priors for group-specific calibration.
These empirical mixture priors were held fixed during the local calculations, thereby incorporating the uncertainty represented by the pilot hyperparameter posterior.
The individually calibrated log learning rates were subsequently shrunk toward the logarithm of the calibrated common rate, as described in Section~\ref{sec:DRW-hie}.
DRW calibration used $B=200$ bootstrap replicates and $S=2000$ retained posterior draws after 1000 burn-in iterations.
Bootstrap weights were generated independently within each group from the scaled Dirichlet distribution.
For these experiments, we set $\kappa=0.01$ and $\gamma=0.30$ in the reference-refit rule of Section~\ref{sec:common-calibration}.
Thus, the reference sample was regenerated when at least $30\%$ of the bootstrap replicates had ESS no greater than $1\%$ of the reference sample size.
We estimated the variance of each calibrated log learning rate using $50$ outer bootstrap replicates and $200$ inner bootstrap replicates.
The finite-difference bandwidth for estimating the derivative of the coverage curve was set to $h=0.1$.
The pilot and final hierarchical fits used 3000 retained MCMC draws after discarding the first 1000 iterations as burn-in.

We compared five methods: individual calibration without shrinkage (IND), individual calibration with shrinkage toward the common rate (SHR), common learning-rate calibration (COM), and fixed learning rates at the lower and upper grid endpoints, $\eta=0.2$ (Fix1) and $\eta=4$ (Fix2).
All methods used the same hyperpriors, loss scale, and number of retained posterior draws for the final hierarchical estimation.
For IND and SHR, the full hierarchy was fitted at the corresponding group-specific learning rates, jointly updating the regression coefficients and hyperparameters.

We evaluated the methods using the root mean squared error (RMSE) of posterior means, the empirical coverage probability (CP), and the average interval length (AIL) of the marginal $95\%$ credible intervals for the group-specific regression coefficients.
Within each replication, these measures were calculated across all groups and both coefficients, and the resulting values were averaged over 200 Monte Carlo replications.
To examine how performance varies with the degree of contamination, we ranked the $m$ groups in increasing order of their contamination probabilities $\omega_i$ and divided them into four subsets of $m/4$ groups each.
These subsets, denoted by Q1--Q4, correspond to progressively higher contamination levels, with Q1 containing the groups with the lowest contamination probabilities and Q4 containing those with the highest.
We calculated CP and AIL separately within each subset by averaging over its groups and both regression coefficients.
The results are reported in Table~\ref{tab:Huber}.
The fixed learning rates show contrasting behavior.
Fix1 produces long intervals, with substantial overcoverage when $m=20$, whereas its overall coverage is close to the nominal level when $m=60$.
In contrast, Fix2 produces short intervals with substantial undercoverage, particularly in groups with higher contamination probabilities.
These results illustrate the limitations of applying a single prespecified learning rate across groups with heterogeneous contamination.
Under IND and SHR, average interval lengths increase more markedly from Q1 to Q4 than under COM or either fixed-rate method, reflecting greater adaptation to the degree of contamination.
However, IND exhibits mild undercoverage, particularly in the most contaminated subset.
This behavior is consistent with instability in individual calibration when the within-group sample size is small.
The proposed shrinkage method improves coverage relative to IND, bringing it close to the nominal $95\%$ level in all four contamination subsets.
Compared with COM, SHR produces shorter intervals while maintaining coverage closer to the nominal level.
Thus, shrinkage provides a more favorable balance between coverage and interval length by stabilizing individual calibration while retaining adaptation to between-group heterogeneity.
These qualitative patterns persist across the different numbers of groups considered.

\begin{table}[htb!]
\centering
\caption{RMSE of posterior means, coverage probabilities (CP; \%), and average interval lengths (AIL) of $95\%$ credible intervals for $\beta_i$, obtained from IND, SHR (proposed), COM, Fix1 ($\eta=0.2$) and Fix2 ($\eta=4$).
Q1--Q4 denote groups divided into four equally sized subsets in increasing order of contamination probability.}
\label{tab:Huber}
\small
\begin{tabular}{ccccccccccccc}
\hline
& & \multicolumn{5}{c}{$m=20$} & & \multicolumn{5}{c}{$m=60$} \\
\cline{3-7}\cline{9-13}
& & IND & SHR & COM & Fix1 & Fix2
  && IND & SHR & COM & Fix1 & Fix2 \\
\hline
RMSE & Overall  & 0.28 & 0.28 & 0.28 & 0.41 & 0.33 &  & 0.28 & 0.27 & 0.32 & 0.44 & 0.31 \\
\hline
 & Q1 & 94.0 & 96.8 & 99.3 & 99.2 & 90.0 &  & 94.2 & 96.1 & 98.1 & 95.6 & 91.2 \\
 & Q2 & 93.4 & 96.4 & 98.8 & 99.3 & 84.6 &  & 92.8 & 95.9 & 97.8 & 95.3 & 85.0 \\
CP (\%) & Q3 & 93.3 & 96.0 & 97.9 & 98.6 & 77.2 &  & 92.0 & 95.2 & 97.4 & 94.7 & 79.7 \\
 & Q4 & 91.6 & 94.2 & 96.2 & 98.7 & 72.2 &  & 91.1 & 94.4 & 96.7 & 94.8 & 72.5 \\
 & Overall  & 93.1 & 95.9 & 98.0 & 98.9 & 81.0 &  & 92.5 & 95.4 & 97.5 & 95.1 & 82.1 \\
\hline
 & Q1 & 0.88 & 0.99 & 1.27 & 2.02 & 0.71 &  & 0.86 & 0.94 & 1.36 & 1.68 & 0.70 \\
 & Q2 & 0.95 & 1.08 & 1.30 & 2.04 & 0.74 &  & 0.90 & 1.02 & 1.39 & 1.70 & 0.73 \\
AIL & Q3 & 1.07 & 1.18 & 1.34 & 2.08 & 0.77 &  & 0.98 & 1.10 & 1.41 & 1.72 & 0.75 \\
 & Q4 & 1.15 & 1.24 & 1.37 & 2.09 & 0.79 &  & 1.08 & 1.18 & 1.43 & 1.73 & 0.78 \\
 & Overall  & 1.01 & 1.12 & 1.32 & 2.06 & 0.75 &  & 0.95 & 1.06 & 1.40 & 1.71 & 0.74 \\
\hline
\end{tabular}
\end{table}

\subsection{Real-data application}
\label{sec:app}

We apply the hierarchical Huber regression model to the longitudinal AIDS data available in the \texttt{catdata} package in R.
The response is the CD4 count divided by 100, and the covariates are time, age, and the CESD depression score.
After removing observations with missing values in these variables, we retain individuals with at least four observations.
The resulting dataset contains $m=315$ individuals and $N=2248$ observations, with individual sample sizes ranging from $4$ to $12$.
The three covariates are standardized, and both the intercept and their regression coefficients are allowed to vary across individuals.
We use the coefficient-wise normal hierarchical prior described above, with diagonal random-coefficient covariance, and set the Huber tuning parameter to $\delta=1.345$.
The residual scale in the Huber loss is fixed at one for all individuals and methods.
We use the same hyperparameter values as in the simulation study in Section~\ref{sec:Huber-sim}.

We compare four approaches: COM (common calibration), SHR (group-specific calibration with shrinkage), and two fixed learning rates at the grid endpoints, $\eta=0.02$ (Fix1) and $\eta=0.8$ (Fix2).
The calibration targets a nominal coverage level of $95\%$, with learning rates evaluated on the grid $\{0.02,0.04,\ldots,0.80\}$.
We use $B=200$ bootstrap replicates for calibration and $R=50$ outer replicates with $B_{\mathrm{var}}=200$ inner replicates for variance estimation.
Calibration uses 2000 retained posterior draws, while the pilot and final hierarchical fits use 3000 retained draws, each following 1000 burn-in iterations.

\begin{figure}[t]
\centering
\includegraphics[width=\textwidth]{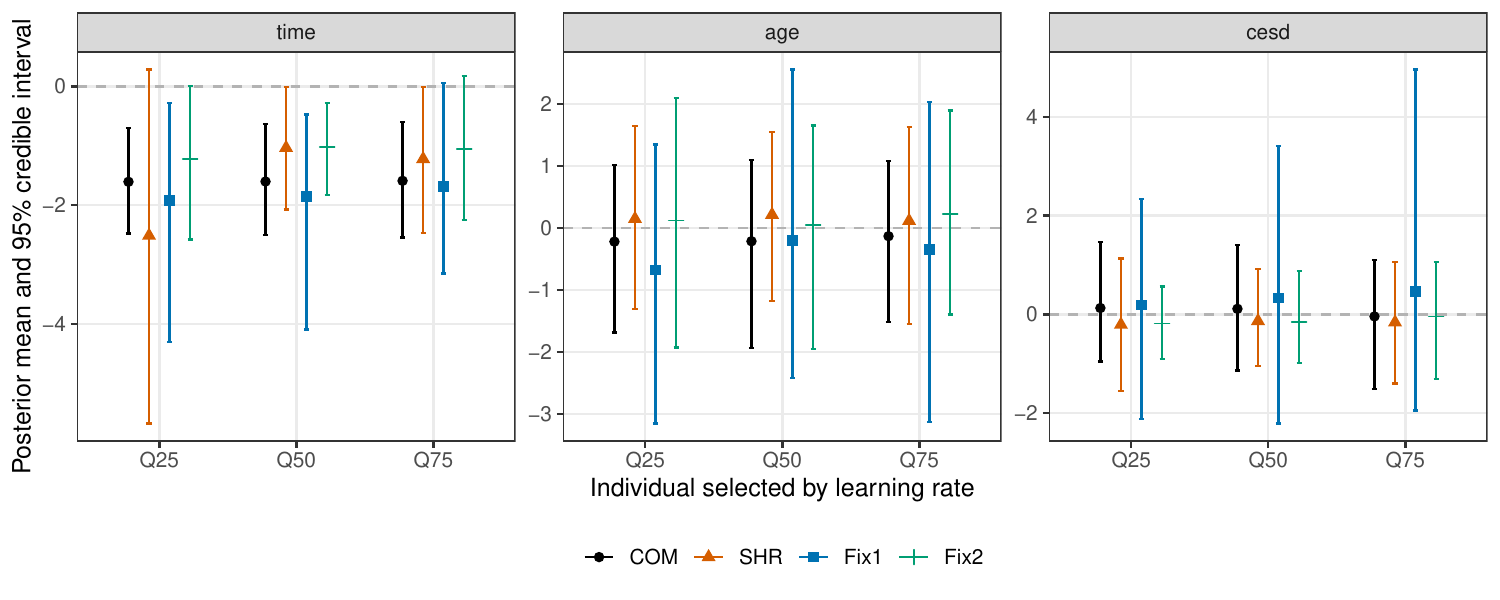}
\caption{Posterior means and 95\% credible intervals for the individual-specific coefficients of time, age, and CESD.
Q25, Q50, and Q75 denote individuals selected near the corresponding empirical quantiles of the estimated learning rates after shrinkage.
The methods are COM, SHR, Fix1 ($\eta=0.02$), and Fix2 ($\eta=0.8$).
The horizontal dashed lines indicate zero.}
\label{fig:huber-ci}
\end{figure}

Figure~\ref{fig:huber-ci} compares posterior means and 95\% credible intervals for three individuals selected near the empirical 25th, 50th, and 75th percentiles of the estimated learning rates after shrinkage.
The same individuals are compared across all methods.
For the time coefficient, SHR produces a substantially wider interval than COM for Q25, whereas its posterior means are less negative than those under COM for Q50 and Q75.
For age and CESD, the intervals include zero under all four methods for the selected individuals, although their locations and widths differ.
Fix1 generally yields wider intervals than Fix2.
The comparison between SHR and COM further shows that individual-specific calibration does not simply expand or contract all intervals uniformly.
Instead, it changes both point estimates and uncertainty assessments in ways that vary across individuals and covariates.

For common learning-rate calibration alone, DRW takes 192 seconds, compared with 5474 seconds for SMC, making DRW approximately 28.6 times faster in this application.
The complete proposed analysis with group-specific learning rates takes 1787 seconds, so the entire group-specific analysis is completed in less time than SMC requires to calibrate a single common learning rate.
Extending SMC to individual calibration and the associated double-bootstrap calculations would require repeatedly targeting many additional posterior distributions and could therefore impose a substantial computational burden.
Although we did not benchmark such an extension, these results demonstrate the computational advantage of the proposed approach for implementing group-specific calibration.

\section{Concluding Remarks}\label{sec:conc}

We proposed a fast calibration method for the learning rate in generalized posterior inference. 
By using Dirichlet-weighted bootstrap perturbations on the loss contributions and combining them with importance reweighting of a small set of reference posterior draws, the method approximates frequentist coverage through a weighted bootstrap criterion while substantially reducing repeated MCMC sampling.
We further extended the framework to hierarchical models, calibrating group-specific learning rates and stabilizing them through a shrinkage step. The numerical examples demonstrated substantial computational savings. In quantile regression, DRW achieved coverage similar to that of existing bootstrap-based calibration methods, while the SVM example produced closely comparable marginal posteriors. In hierarchical Huber regression, shrinkage improved coverage relative to individual calibration without shrinkage.

Several directions merit further study.
First, extending the fixed-learning-rate justification to a formal coverage guarantee at the data-dependent selected rate would require uniform control of the distributional approximations and numerical errors.
Second, the current criterion targets average marginal coverage; calibration of simultaneous credible regions is an important extension.
Third, further analysis of group-specific calibration and shrinkage would clarify their effects on coverage when within-group sample sizes are small.
Finally, adaptive selection of reference distributions and ESS thresholds could improve computational efficiency.

\section*{Acknowledgements}
We would like to thank Masahiro Tanaka for useful discussions and comments. 
This work is partially supported by Japan Society for the Promotion of Science (JSPS) KAKENHI Grant Numbers 24K21420 and 25H00546.

\bibliographystyle{chicago}
\bibliography{ref}

@article{bissiri2016general,
  title={A general framework for updating belief distributions},
  author={Bissiri, Pier Giovanni and Holmes, Chris and Walker, Stephen G.},
  journal={Journal of the Royal Statistical Society: Series B},
  volume={78},
  number={5},
  pages={1103--1130},
  year={2016}
}

@inproceedings{grunwald2012safe,
  title={The safe Bayesian: learning the learning rate via the mixability gap},
  author={Gr{\"u}nwald, Peter},
  booktitle={International Conference on Algorithmic Learning Theory},
  pages={169--183},
  year={2012},
  organization={Springer}
}

@article{grunwald2020inconsistency,
  title={Inconsistency of Bayesian inference for misspecified models},
  author={Gr{\"u}nwald, Peter and van Ommen, Thijs},
  journal={Entropy},
  volume={22},
  number={10},
  pages={1121},
  year={2020}
}

@article{holmes2017assigning,
  title={Assigning a value to a power likelihood in a general {B}ayesian model},
  author={Holmes, C. C. and Walker, S. G.},
  journal={Biometrika}, volume={104}, number={2}, pages={497--503}, year={2017}}

@article{huber1973robust,
  title={Robust regression: asymptotics, conjectures and Monte Carlo},
  author={Huber, Peter J},
  journal={The annals of statistics},
  pages={799--821},
  year={1973},
  publisher={JSTOR}
}

@article{jiang2008gibbs,
  title={Gibbs Posterior for Variable Selection in High-Dimensional Classification and Data Mining},
  author={Jiang, Wenxin and Tanner, Martin A},
  journal={The Annals of Statistics},
  pages={2207--2231},
  year={2008},
  publisher={JSTOR}
}

@article{kawakami2023approximate,
  title={Approximate Gibbs sampler for Bayesian huberized lasso},
  author={Kawakami, Jun and Hashimoto, Shintaro},
  journal={Journal of Statistical Computation and Simulation},
  volume={93},
  number={1},
  pages={128--162},
  year={2023},
  publisher={Taylor \& Francis}
}

@article{kozumi2011gibbs,
  title={Gibbs sampling methods for Bayesian quantile regression},
  author={Kozumi, Hideo and Kobayashi, Genya},
  journal={Journal of statistical computation and simulation},
  volume={81},
  number={11},
  pages={1565--1578},
  year={2011},
  publisher={Taylor \& Francis}
}

@article{li2010adjusted,
  title={An adjusted maximum likelihood method for solving small area estimation problems},
  author={Li, Huilin and Lahiri, Partha},
  journal={Journal of multivariate analysis},
  volume={101},
  number={4},
  pages={882--892},
  year={2010},
  publisher={Elsevier}
}

@article{lyddon2019general,
  title={General Bayesian updating and the loss-likelihood bootstrap},
  author={Lyddon, Simon and Holmes, Chris and Walker, Stephen},
  journal={Biometrika},
  volume={106},
  number={2},
  pages={465--478},
  year={2019}
}

@article{miller2021asymptotic,
  title={Asymptotic normality, concentration, and coverage of generalized posteriors},
  author={Miller, Jeffrey W.},
  journal={Journal of Machine Learning Research},
  volume={22},
  number={168},
  pages={1--53},
  year={2021}
}

@article{matsubara2024generalized,
  title={Generalized {B}ayesian inference for discrete intractable likelihood},
  author={Matsubara, Takuo and Knoblauch, Jeremias and Briol, Fran{\c c}ois-Xavier and Oates, Chris J.},
  journal={Journal of the American Statistical Association},
  volume={119}, 
  number={547}, 
  pages={2345--2355}, 
  year={2024}
}

@article{newton1994approximate,
  title={Approximate Bayesian inference with the weighted likelihood bootstrap},
  author={Newton, Michael A. and Raftery, Adrian E.},
  journal={Journal of the Royal Statistical Society: Series B},
  volume={56},
  number={1},
  pages={3--48},
  year={1994}
}

@article{nott2024bayesian,
  title={Bayesian Inference for Misspecified Generative Models},
  author={Nott, David J and Drovandi, Christopher and Frazier, David T},
  journal={Annual Review of Statistics and Its Application},
  volume={11},
  number={1},
  pages={179--202},
  year={2024},
  publisher={Annual Reviews}
}

@article{onizuka2026learning,
  title={Learning rate selection via weighted {F}isher divergence},
  author={Onizuka, Takahiro},
  journal={arXiv preprint arXiv:2606.26478}, year={2026}
}

@article{praestgaard1993exchangeably,
  title={Exchangeably weighted bootstraps of the general empirical process},
  author={Pr{\ae}stgaard, Jens and Wellner, Jon A},
  journal={The Annals of Probability},
  pages={2053--2086},
  year={1993},
  publisher={JSTOR}
}

@article{polson2011data,
  title={Data augmentation for support vector machines},
  author={Polson, Nicholas G and Scott, Steven L},
  journal={Bayesian Analysis},
  volume={6},
  pages={1--24},
  year={2011}
}

@article{rubin1981bayesian,
  title={The Bayesian bootstrap},
  author={Rubin, Donald B.},
  journal={The Annals of Statistics},
  volume={9},
  number={1},
  pages={130--134},
  year={1981}
}

@article{syring2019calibrating,
  title={Calibrating general posterior credible regions},
  author={Syring, Nicholas and Martin, Ryan},
  journal={Biometrika},
  volume={106},
  number={2},
  pages={479--486},
  year={2019}
}

@inproceedings{tanaka2024generalized,
  title={Generalized posterior calibration via sequential monte carlo sampler},
  author={Tanaka, Masahiro},
  booktitle={Proceedings of the 2024 6th Asia Conference on Machine Learning and Computing},
  pages={62--68},
  year={2024}
}

@article{tamano2025location,
  title={Location--scale calibration for generalized posterior},
  author={Tamano, Shu and Tomo, Yui},
  journal={arXiv preprint arXiv:2511.15320}, year={2025}
}

@article{wu2023comparison,
  title={A comparison of learning rate selection methods in generalized Bayesian inference},
  author={Wu, Pei-Shien and Martin, Ryan},
  journal={Bayesian Analysis},
  volume={18},
  number={1},
  pages={105--132},
  year={2023},
  publisher={International Society for Bayesian Analysis}
}

\end{document}